# Propagation of Relativistic Shock Wave Induced by Laser Spark in Quiescent Air

**Awanish Pratap Singh**        **Upasana P. Padhi**        **Ratan Joarder**

Aerospace Engineering Department  Aerospace Engineering Department  Aerospace Engineering Department
IIT Kharagpur, 721302             IIT Kharagpur, 721302             IIT Kharagpur, 721302
awanishsingh009@iitkgp.ac.in      upasana0904@gmail.com             jratan@aero.iitkgp.ac.in

**Abstract**
Laser-induced breakdown has shown many potential applications in the various field of science and engineering. As the breakdown occurs in gas/aerosol, a rapid hydrodynamic expansion as shock (blast) wave initiated from the deposition location. The nature of the shock wave is one of the controlling factors in many physical processes; unfortunately, it's nature has still not been clearly understood. In this study, an error was found during the calculation of shock wave properties with the classical non-relativistic approach. The error in calculation was due to the initial relativistic propagation of high-temperature plasma. Initially, the plasma and shock wave travel together with higher acceleration up to the point of inflation, and in later time shock wave dissociate itself from the plasma. However, this accelerating effect is neglected in the earlier studies. To address the spontaneous accelerating and deaccelerating nature of the shock wave, the theoretical details of the relativistic approach of shock wave propagation is presented.

Keywords: *Laser-induced breakdown; Relativistic shock; Shock-wave; Hydrodynamic expansion.*

## I. INTRODUCTION

Laser-induced breakdown (LIB) of air has recently attracted considerable interest due to its promising applications. For example, the interest in lasers for ignition and combustion under different conditions [1,2], aerodynamic drag reduction [3–5], lift and moment control as well as local flow-field alternation [6], propulsion system [7], and laser thrusters for space propulsion [8]. Extensive research has also been performed to understand the process of energy deposition by laser, blast-wave propagation, and expansion of the plasma kernel. When the LIB takes place in gas/aerosol, it causes intense heating of nearby free electrons and heavy particles (ion, atoms, molecules etc.) leads to rapid hydrodynamic expansion in the form of an ellipsoidal shock wave up to first 3-4 $\mu$s (asymmetrical expansion) and in later time it converts in to a spherical shock wave (symmetrical expansion). The nature of the blast-wave is one of the controlling factors in the formation of the flame kernel and the flame propagation process. Therefore, the behaviour of the blast-wave should be clearly understood to understand the other associated physical phenomena.

According to the available literatures [1], the initial velocity of the shock can be greater than $10^7$ cm/s, which is only three-order less than speed of light, whereas the velocity of electrons and heavy particles (ion, atoms, molecule etc.) during the collision and initial expansion can be even higher (around $10^9$ cm/s) during multiphoton absorption and inverse-bremsstrahlung (IB) process. In this study, the error associated with calculations of shock wave propagation are presented. Two different methods are used to calculate the shock velocity, where both methods showed great a discrepancy between each other. The reason for the discrepancy was identified as the faulty calculation of very fast shock wave expansion with the non-relativistic approach. After the formulation of shock wave theory by Hugoniot in 1887, it has been widely used in various fields of science and engineering due to its consistency in the prediction of the physical phenomena. Since then three different and independent mechanisms have been proposed to account for laser-induced shock wave phenomenon, i.e. radiation supported shock wave, breakdown wave and radiation transport wave. Later, an additional viewpoint has been suggested when the laser pulse is ended, i.e. blast wave theory. However, these theories failed to correctly predict the fast-hydrodynamic expansion, which is also reported in many literatures [1,2]. Therefore, it is necessary to use the relativistic approach for the calculation for shock wave propagation to avoid the obtained errors. To address the above issues, this paper highlights, the relativistic Rankine-Hugoniot relations, which can be used to calculate the properties of shock wave induced by laser spark or any kind of intense blast.

In section II we briefly describe the experimental details, the equipment used in the experiments, and the theoretical details related to relativistic approach. In section III, the characteristics of the blast wave after a laser-induced breakdown are presented, and a few important properties of relativistic blast waves which make them amenable to approximation are discussed. Section IV is the conclusion.

## II. METHODOLOGY

### A. Experimental Details

A schematic of the experimental setup consisting of a Q-switched Nd: YAG laser (Quantel EVG00200) with dual cavities was used to study the propagation of shock wave induced by laser spark is shown in Fig. 1. The laser was capable of producing two pulses at a very short interval (in nanoseconds) in the range of 10-200 mJ per pulse at a wavelength of 532 nm with a pulse duration of 7 ns. The laser beam was expanded and then focused in the quiescent air at atmospheric pressure to induce breakdown by using 50.8 mm diameter achromatic doublet lens ($f$ = 150 mm). To precisely capture the laser-induced breakdown phenomena, a reflect-type high-speed schlieren imaging system was used for the flow field visualization. Schlieren images were recorded with a high-speed ICCD camera (4 Quick E, Stanford computer optics), which was equipped with a Nikkor 55-200 mm f/4-5.6G ED VR II lens. The camera was operated at an exposure time of 100 ns, and the spatial resolution (78.125 μm/pixel) was determined by using a calibration target. A digital delay generator (BNC-745) of femtosecond accuracy was used to control the operation time of the camera and the laser system.

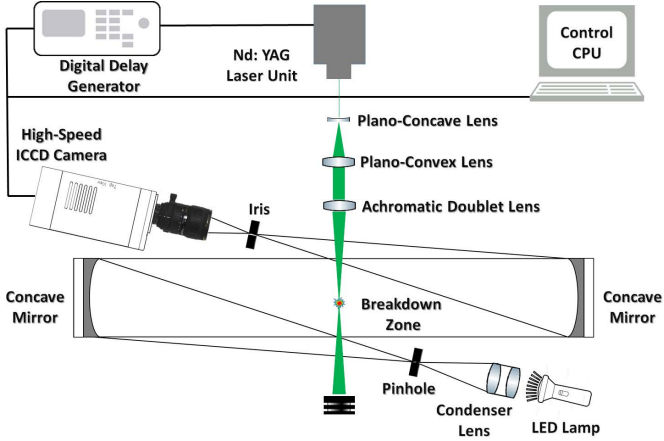

**Figure 1: Schematic of the experimental setup.**

### B. Theoretical Details

The properties of shock wave induced by laser spark were calculated in many literatures from the classical Rankine–Hugoniot relationships [2,9]. The present study discusses the need of a relativistic approach to calculate the shock properties. The theoretical foundation of relativistic shock waves is based on relativistic hydrodynamics Landau and Lifshitz [10] and was first analysed by Taub [11]. Eliezer et al. [12] studied the theoretical and experimental aspect of laser-induced shock waves for the laser intensities $10^{12}$ W cm$^{-2}$ < $I_L$ < $10^{16}$ W cm$^{-2}$ and nanoseconds pulse duration.

Newton's Second Law, which we expressed by the equation, $F = d(mv)/dt$ was stated with tacit assumption that mass ($m$) is constant. However, now we know that it is not true and the mass of a body increases with velocity [13]. Einstein's corrected the formula $m$ has the value,

$$m = m_1 / \sqrt{1 - v_s^2 / c^2} \quad (1)$$

Here, in this case, $m_1$ is the mass at rest of either electron or heavy particles in the plasma, $v_s$ is the shock velocity and $c$ is the speed of light. Now, the momentum is still given by $mv$, but when we use the new $m$ it becomes

$$\boldsymbol{p} = m\boldsymbol{v} = m_1\boldsymbol{v} / \sqrt{1 - v_s^2 / c^2} \quad (2)$$

Now let's see why the relativistic approach is needed for very fast expansion mechanism. During photoionization or IB process the electrons are colliding very fast with the heavy particles (approximately within nanoseconds). If before ionization the mass of the atom was $m_0$ at rest, and the mass of ion and electron was $m_i$ and $m_e$ respectively. Then after the ionization, when both ion and electron is moving very fast, and if we add both the mass together, the mass will not remain equal to the mass of an atom at rest, which is ($m_i + m_e \neq m_0$). It is because the mass of ion and especially of electron might have enhanced over the mass when they were together and standing still. In the similar manner, the Lorentz transformation can be used for the velocity transformation and other related parameters for the behavior of relativistic hydrodynamic expansion. The energy-momentum 4-tensor $T_{\mu\nu}$ governs the behaviour of relativistic hydrodynamic expansion is given by,

$$T_{\mu\nu} = hU_\mu U_\nu + Pg_{\mu\nu} \quad (3)$$

In this equation, $h = e + P$ is the enthalpy, $e$ is the mass-energy density of the fluid in the comoving frame, $P$ is the pressure, $U_\mu = (\gamma c, \gamma v_1, \gamma v_2, \gamma v_3)$ is the fluid 4-velocity, $\gamma = (1 - \beta^2)^{-1/2}$ is the fluid Lorentz factor, $\beta = v/c$ is the velocity of the fluid in units of $c$, where $c$ is the speed of light. The Minkowski metric tensor $g_{\mu\nu}$ : $g_{00} = -1$, $g_{11} = g_{22} = g_{33} = 1$, $g_{\mu\nu} = 0$ if $\mu \neq \nu$, for $\mu, \nu$ = 0, 1, 2, 3. The following equations explicitly give the energy-momentum conservation, the particle number conservation and the equation of states (EOS),

$$\frac{\partial T_\mu^\nu}{\partial x^\nu} \equiv \partial_\nu T_\mu^\nu = 0 \; for \; \mu = 0, 1, 2, 3. \quad (4)$$

$$\frac{\partial (nU^\mu)}{\partial x^\mu} \equiv \partial_\mu (nU^\mu) = 0 \quad (5)$$

$$P = P(e, n) \quad (6)$$

For solving the equation of state, the ideal gas equation $e = \rho c^2 + P/(\Gamma - 1)$ is taken into consideration to calculate the shock-wave properties. After solving the equation 4, 5 and 6 for the ideal gas, the following solution can be obtained in the laboratory frame of reference.

$$\frac{u_s}{c} = \left[\frac{(e_1 + P_0)(P_1 - P_0)}{(e_1 - e_0)(e_0 + P_1)}\right]^{1/2} \quad (7)$$

$$\frac{u_p}{c} = \left[\frac{(e_1 - e_0)(P_1 - P_0)}{(e_1 + P_0)(e_0 + P_1)}\right]^{1/2} \quad (8)$$

$$\frac{(e_1 + P_1)^2}{\rho_1^2} - \frac{(e_0 + P_0)^2}{\rho_0^2} = (P_1 - P_0)\left[\frac{(e_1 + P_1)}{\rho_1} + \frac{(e_0 + P_0)}{\rho_0}\right] \quad (9)$$

$$e_j = \rho_j c^2 + \frac{P_j}{\Gamma - 1}; \quad j = 0, 1. \quad (10)$$

Where $u_s$ is the shock wave velocity, $u_p$ is the particle flow velocity, $\rho$ is the mass density, and $\Gamma$ is the specific heat ratio. The subscripts 0 and 1 denote the domains before and after the shock arrival. If the $P$ is much smaller than $\rho c^2$, then, $u_s/c \ll 1$. Therefore, above equations (7-10), yield to classical Rankine-Hugoniot equations. The numerical detail of relativistic approach is beyond the scope of this paper but will be accommodated in future studies.

## III. RESULTS AND DISCUSSION

The sequence of physical events from energy deposition to shock wave propagation at different time after the energy deposition is shown in Fig. 2. The energy deposited during the air-breakdown is measured as 42 mJ out of 50 mJ pulse energy. The high-speed schlieren imaging technique is used to record the shock wave propagation after the air breakdown. The shock wave diameters are measured with an in-house Matlab algorithm with an accuracy of 100 μm and therefore no error bars are displayed in Fig. 3. Fig. 3(a) shows the measured expansion trajectory, and shock speed with two different methods is shown in Fig. 3(b). For better visualization, each image in Fig. 2 is enlarged and rescaled using an in-house Matlab algorithm with size $y \times x$ mm.

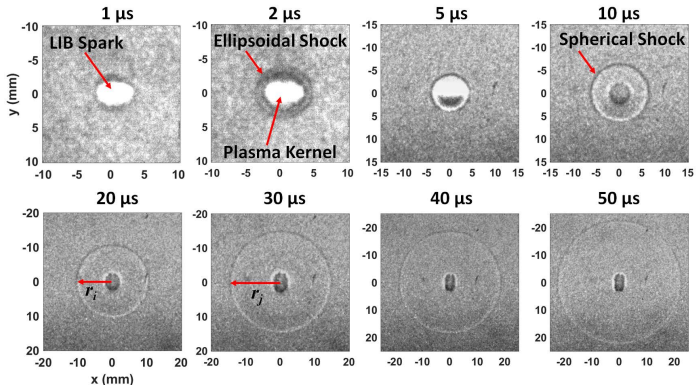

Figure 2: Schlieren images of the shock wave propagation in quiescent air.

In one of the initial study by Ramsden and Savic [14], it was observed that both plasma front and shock wave propagate with time dependency of about $t^{0.6}$. Similar observation of the time dependency of about $t^{0.6}$ was also reported in conventional spark ignition [15]. In the present experimental study, the time dependency of approximately $t^{0.68}$ was observed, which is shown in Fig. 3(a). The classical blast wave theory, which is valid after energy transfer to plasma has ended, shows the dependency of about $t^{0.4}$. The difference in time dependency may arise due to different energy value, pulse duration, optical setup, etc. However, the key reason for the difference in time dependency is due to the measurement of spontaneous deaccelerating nature of the shock wave with classical methods. The reason became evident when the experimental data was splitted from 1-50 μs (Fig. 3 (a)) to 1-10 μs and 10-50 μs, they both showed the best fit for different time dependency of approximately $t^{0.56}$ and $t^{0.79}$ respectively. It is also evident from here that the time dependency is scale dependent. Therefore, the nature of the shock wave cannot be predicted with the classical methods. The Mach lines (Ma = 1, 2, 3, 4) are plotted in Fig. 3 (a) to show the deaccelerating nature of the shock wave. It can be seen that the classical blast wave theory and other classical methods show inconstancy with the experimental results. Therefore, these methods are not adequate and suitable to calculate the shock properties and inverse-bremsstrahlung coefficients, for the deaccelerating nature of the shock wave. Therefore, to address this problem, the discussion of relativistic approach is required.

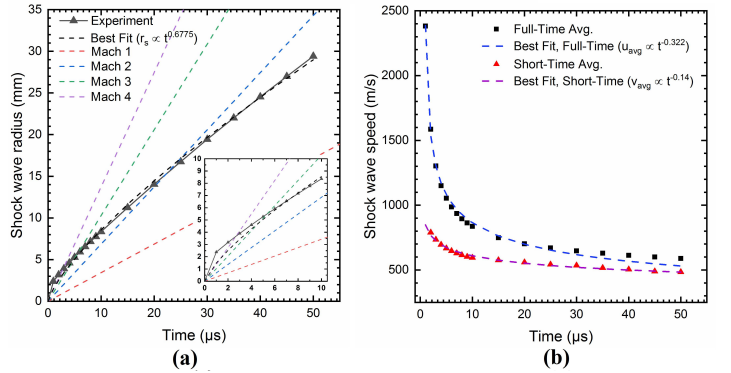

Figure 3: After the energy deposition of 50 mJ per pulse (a) variation of shock radius (b) shock speed at the different instant of time.

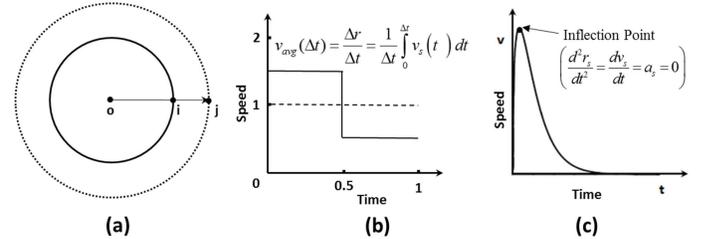

Figure 4: Representation of shock wave (a) at different position, (b) velocity by two different approaches, (c) inflection point.

Chiu [16] in his study shows that for the stronger shock the relativistic theory as shown in equation (7-9) deviates from classical Rankine-Hugoniot theory because the relativistic theory predicts higher particle density for a given pressure condition. He also emphasized that if the initial state is at very high temperature, the relativistic theory will deviate from the classical theory. The similar effect of deviation is presented in Fig. 3(b) when two different approaches were used to calculate the shock wave velocity. The reason for the error in the calculation arises in the present as well as in previous studies is presented in the subsequent discussion.

In Fig. 4(a) three points (*o, i, j*) are shown. Point *o* is the origin, *i* and *j* are the position of shock at the different time interval after the energy deposition, which is also shown in Fig. 2 (20 μs, 30 μs). In Fig. 3(b) velocity of the shock is calculated by two different classical methods, i.e. full time and short time approach. The similar method is used in Fig. 4(b) to explain the reason for the error in calculation. In Fig. 4(b) dash line is the speed calculated by the full-time approach and continuous line by the short-time approach. For the full-time approach, the following conditions are used for the calculation $\Delta r = r_j - r_i$ for $r_i = 0$ and $\Delta t = t_j - t_i$ for $t_i = 0$. Now let us take an example to understand the problem. Suppose that if the shock wave travels one unit of space in one unit of time, the speed by the full method will be one unit as shown in Fig. 4 (b) by the dashed line. If we discretise the time in half as shown in Fig. 4(b) by an unbroken line, then let's say shock wave travels 0.75 unit of space in 0.5 unit of time then speed will be 1.5 unit for the first half, and for another half the speed will be 0.5 unit because shock has travelled only 0.25 unit of space in 0.5 unit of time. It is clearly observed from Fig. 4(b) that at the middle of the total observed time both the method will intersect each other, but in Fig 3(b) it is not observed anywhere. However, Fig. 4(b) is not the exact representation of Fig. 3(b), but still there will an intersection between two approaches but at a different unit of time (slightly later). The reason for the error in calculation is due to consideration of non-relativistic approach and the assumption that shock-wave is decelerating from the beginning of plasma formation. Whereas in physical reality, the shock wave first accelerates with the very high-temperature plasma, and after reaching the point of inflection (point of maximum velocity), it starts decelerating, as shown in Fig. 4(c). The above discussion shows that the non-relativistic approach is not sufficient to study the phenomena mentioned above. This paper only shows the possibility for the relativistic approach and the reason for its necessity. The implication of theoretical details is beyond the scope of this paper but will be used in the future study.

## IV. CONCLUSIONS

After the breakdown, the shock wave first accelerates with the very high-temperature plasma, and after reaching the point of inflection, it starts deaccelerating. The initial speed and temperature are so high that the classical Rankine-Hugoniot theory failed to predict the exact physical phenomena. In this study, the reason for failure when compared with the experimental result is explained. The reason for the discrepancy was identified as the faulty calculation of very fast shock wave expansion with the non-relativistic approach. Therefore, this study suggests the necessity of a relativistic approach with theoretical details, which can be used in the future study of the propagation of shock(blast) wave induced by laser spark. The relativistic theory predicts higher particle density for a given pressure condition during initial expansion of high-temperature plasma.